\documentclass[12pt]{article} 
\usepackage{graphicx}
\begin{document}

\title{Improved Component Predictions of Batting Measures}
\author{Jim Albert \\Department of Mathematics and Statistics \\ Bowling Green State University}

\maketitle
\begin{abstract}
Standard measures of batting performance such as a batting average and an on-base percentage can be decomposed into component rates such as strikeout rates and home run rates.  The likelihood of hitting data for a group of players can be expressed as a product of likelihoods of the component probabilities and this motivates the use of random effects models to estimate the groups of component rates.  This methodology leads to accurate estimates at hitting probabilities and good predictions of performance for following seasons.  This approach is also illustrated for on-base probabilities and FIP abilities of pitchers.
\end{abstract}

\section{Introduction}

Efron and Morris (1975) demonstrated the benefit of simultaneous estimation using a simple example of using the batting outcomes of 18 players in the first 45 at-bats in the  1970 season to predict their batting average for the remainder of the season.  Essentially, improved batting estimates shrink the observed averages towards the average batting average of all players.  One common way of achieving this shrinkage is by means of a random effects model where the players' underlying probabilities are assumed to come from a common distribution, and the parameters of this ``random effects'' distribution are assigned vague prior distributions.

In modern sabermetrics research, a batting average is not perceived to be a valuable measure of batting performance.  One issue is that the batting average assigns each possible hit the same value, and it does not incorporate in-play events such as walks that are beneficial for the general goal of scoring runs.  Another concern is that the batting average is a convoluted measure that combines different batting abilities such as not striking out, hitting a home run, and getting a hit on a ball placed in-play.  Indeed, it is difficult to really say what it means for a batter to ``hit for average''.  Similarly, an on-base percentage does not directly communicate a batter's different performances in drawing walks or getting base hits.

A deeper concern about a batting average is that chance plays a large role in the variability of player batting averages, or the variability of a player's batting average over seasons.  Albert (2004) uses a beta-binomial random effects model to demonstrate this point.  If a group of players have 500 at-bats, then approximately half the variability in the players' batting average is due to chance (binomial) variation the remaining half is due to variability in the underlying player's hitting probabilities.  In contrast, other batting rates are less affected by chance.  For example, only a small percentage of players observed home run rates are influenced by chance -- much of the variability is due to the differences in the batters' home run abilities.

The role of chance has received recent attention to the development of FIP (fielding independent pitching) measures.  McCracken (2001) made the surprising observation that pitchers had little control of the outcomes of balls that were put in-play.  One conclusion from this observation is that the BABIP, batting average on balls put in-play, is largely influenced by luck or binomial variation, and the FIP measure is based on outcomes such as strikeouts, walks, and home runs that are largely under the pitcher's control.

Following Bickel (2004), Albert (2004) illustrated the decomposition of a batting average into different components and discussed the luck/skill aspect of different batting rates. In this paper, similar decompositions are used to develop more accurate predictions of a collection of batting averages.  Essentially, the main idea is to first represent a hitting probability in terms of component probabilities, estimate groups of component probabilities by means of random effects models, and use these component probability estimates to obtain accurate estimates of the hitting probabilities.  Sections 3, 4, 5 illustrate the general ideas for the problem of simultaneously estimating a collection of ``batting average'' probabilities and Section 8 demonstrates the usefulness of this scheme in predicting batting averages for a following season. Section 7 illustrates this plan for estimating on-base probabilities.  Section 8 gives a historical perspective on how the different component hitting rates have changed over time, and Section 9 illustrates the use of this perspective in understanding the career trajectories of hitters and pitchers. The FIP measure is shown in Section 10 as a function of particular hitting rates and this representation is used to develop useful estimates of pitcher FIP abilities.  Section 11 concludes by describing several modeling extensions of this approach.

\section{Related Literature}

Since Efron and Morris (1975), there is a body of work finding improved measures of performance in baseball.  Tango et al (2007) discuss the general idea of estimating a player's true talent level by adjusting his past performance towards the performance of a group of similar players and the appendix gives the familiar normal likelihood/normal prior algorithm for performing this adjustment.  Brown (2008), McShane et al  (2011), Neal et al (2010), and Null (2009) propose different ``shrinking-type'' methods for estimating batting abilities for Major League batters.  Similar types of methods are proposed by Albert (2006), Piette and James (2012), and Piette et al (2010) for estimating pitching and fielding metrics.  Albert (2002) and Piette et al (2012) focus on the problem of simulataneously estimating player hitting and fielding trajectories.

Albert (2004), Bickel (2004) and Bickel and Stotz (2003) describe  decomposition of batting averages.  Baumer (2008) performs a similar decomposition of batting average ($BA$) and on-base percentage ($OBP$) with the intent of showing mathematically that $BA$ is more sensitive than   $OBP$ to the batting average on balls in-play.

\section{Decomposition of a Batting Average.}

The basic decomposition of a batting average is illustrated in Figure 1.  Suppose one divides all at-bats into strikeouts (SO) and not-strikeouts (Not SO).  Of the AB that are not strikeouts, we divide into the home runs (HR) and the balls that are put ``in-play''.  Finally, we divide the balls in-play into the in-play hits (HIP) and the in-play outs (OIP).

\begin{figure}[h]
\begin{center}
 \includegraphics[scale=0.4]{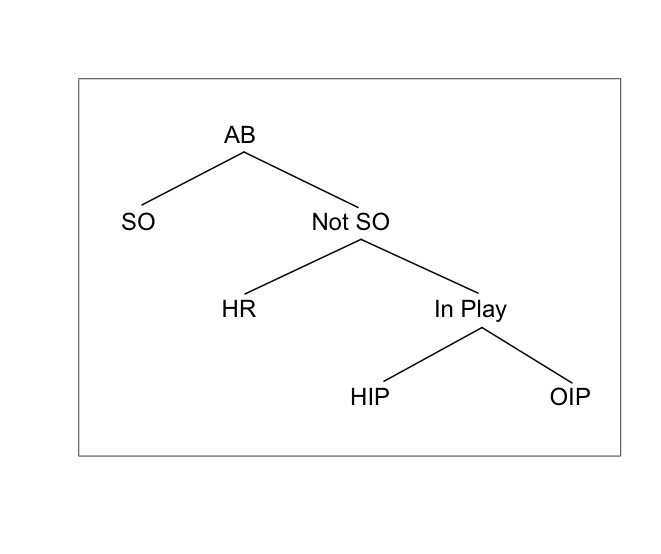}
\end{center}
\caption{Breakdown of an at-bat.}
\end{figure}

This representation leads to a decomposition of the batting average $H / AB$.  We first write the proportion of hits as the proportion of AB that are not strikeouts times the proportion of hits among the non-strikeouts.
$$
\frac{H}{AB} = \left(1 - \frac{SO}{AB}\right) \times \frac{H}{AB - SO}.
$$
Continuing, if we breakdown these $AB - SO$ opportunities by $HR$, then we write the hit proportion as the proportion of non-strikeouts that are home runs plus the proportion of non-strikeouts that are singles, doubles, or triples $(H - HR)$.
$$
\frac{H}{AB - SO} = \frac{HR}{AB - SO} + \frac{H - HR}{AB-SO}
$$
Finally, we write the proportion of non-strikeouts that are singles, doubles, or triples as the proportion of non-strikeouts that are not home runs times the proportion of balls in-play ($AB - SO - HR$) that are hits.
$$
 \frac{H - HR}{AB-SO} = \left(1 - \frac{HR}{AB-SO}\right)  \times  \frac{H - HR}{AB-SO-HR}. 
$$

Putting it all together, we have the following representation of a batting average $BA = H / AB$:
$$
BA = (1 - SO.Rate) \times \left( HR.Rate + (1 - HR.Rate) \times BABIP\right),
$$
where the relevant rates are:
\begin{itemize}
\item the strikeout rate $$ SO.Rate = \frac{SO}{AB} $$
\item the home run rate $$ HR.Rate = \frac{HR}{AB - SO} $$
\item the batting average on balls in play rate $$BABIP = \frac{H - HR}{AB - SO - HR}$$.
\end{itemize}
Instead of simply recording hits and outs, we are regarding the outcomes of an at-bat as  multinomial data with the four outcomes SO, HR, HIP, and OIP.

\section{Multinomial Data and Likelihood}

The decomposition of a batting average leads to a multinomial sampling model for the hitting outcomes, and the multinomial sampling leads to an attractive representation of the likelihood of the underlying probabilities.

There are four outcomes of an at-bat:  SO, HR, HIP, and OIP (strikeout, home run, hit-in-play, and out-in-play).  Let $p_{SO}$ denote the probability that an at-bat results in a strikeout, let $p_{HR}$ denote the probability a non-strikeout results in a home run, and $p_{HIP}$ denotes the probability that a ball-in-play (not SO or HR) results in an in-play hit.
If this player has $n$ at-bats, the vector of counts of SO, HR, HIP, and OIP is  multinomial with corresponding probabilities $p_{SO}, (1 - p_{SO}) p_{HR}, (1 - p_{SO}) (1 - p_{HR}) p_{HIP}$, and $(1 - p_{SO}) (1 - p_{HR}) (1 - p_{HIP}).$  These expressions are analogous to the breakdowns of the hitting rates.  For example, the probability a hitter gets a home run in an at-bat is equal to the probability that the person does not strike out ($1 - p_{SO}$), times the probability that the hitter gets a home run among all the non-strikeouts ($p_{HR}$).  Likewise, the probability a batter gets a hit in-play is the probability he does not strikeout times the probability he does not get a home run in a non-strikeout times the probability he gets a hit in a ball put into play.

Denote the multinomial counts for a particular player as $(y_{SO}, y_{HR}, y_{HIP}, n - y_{SO} - y_{HR} - y_{HIP})$ where $n$ is the total number of at-bats.   The likelihood of the associated probabilities is given by
\begin{eqnarray*}
L & = & p_{SO} ^ {y_{SO}} \times \left( (1 - p_{SO})p_{HR} \right)^{y_{HR}} \times
     \left( (1 - p_{SO})(1 - p_{HR}) p_{HIP} \right)^{y_{HIP}} \\
      & & \times  \left( (1 - p_{SO})(1 - p_{HR}) (1 - p_{HIP}) \right)^{n - y_{SO} - y_{HR} - y_{HIP}} .\\
\end{eqnarray*}
With some rearrangement of terms, one can show that the likelihood has a convenient factorization:
\begin{eqnarray*}
 L     & = & \left[p_{SO}^{y_{SO}} (1 - p_{SO})^{n - y_{SO}}\right] \times
               \left[p_{HR}^{y_{HR}} (1 - p_{HR})^{n - y_{SO} - y_{HR}}\right]\\
    & & \times \, \, \left[p_{HIP}^{y_{HIP}} (1 - p_{HIP})^{n - y_{SO} - y_{HR} - y_{HIP}}\right] \\
    & = & L_1 \times L_2 \times L_3
\end{eqnarray*}

From the above representation, we see
\begin{itemize}
\item $L_1$ is the likelihood for a binomial($n, p_{SO}$) distribution
\item $L_2$ is the likelihood for a binomial($n - y_{SO}, p_{HR}$) distribution
\item $L_3$ is the likelihood for a binomial($n - y_{SO} - y_{HR}, p_{HIP}$) distribution
\end{itemize}

Above we consider the multinomial likelihood for a single player, when in reality we have $N$ hitters with unique hitting probabilities.
For the $j$th player, we have associated probabilities $p_{SO}^j, p_{HR}^j, p_{HIP}^j$.  Following the same factorization, it is straightforward to show that the likelihood of the vectors of probabilities ${\bf p_{SO}} = \{ p_{SO}^j\}, {\bf p_{HR}} = \{p_{HR}^j\}$, and  ${\bf p_{HIP}} = \{p_{HIP}^j\}$ is given by
\begin{eqnarray*}
L({\bf p_{SO}}, {\bf p_{HR}}, {\bf p_{HIP}}) & =&  \prod_{j=1}^N ( L_1^j L_2^j L_3^j) \\
& = & \prod_{j=1}^N L_1^j \times \prod_{j=1}^N L_2^j \times \prod_{j=1}^N L_3^j
\end{eqnarray*}

\section{Exchangeable Modeling}

\subsection{The Prior}

The factorization of the likelihood motivates an attractive way of simultaneously estimating the multinomial probabilities for all players.  Suppose that one's prior belief about the vectors ${\bf p_{SO}}$, ${\bf p_{HR}}$, and ${\bf p_{HIP}}$ are independent, and we represent each set of probabilities by an exchangeable model represented by a multilevel prior structure.

In particular,  suppose that the strikeout probabilities are believed to be exchangeable.  One way of representing this belief is by the following mixture of betas model.
\begin{itemize}
\item $p_{SO}^1, ..., p_{SO}^N$ are independent $Beta(K_{SO}, \eta_{SO})$, where a $Beta(K, \eta)$ density has the form
$$
g(p) = \frac{1}{B(K \eta, K(1- \eta))} p^{K \eta - 1} (1 - p)^{K (1 - \eta) - 1}, \, \, 0 < p < 1,
$$
and $B(a, b)$ is the beta function.  The parameter $\eta$ is the prior mean and $K$ is a ``precision'' parameter in the sense that the prior variance $\eta(1 - \eta) / (K + 1)$ is a decreasing function of $K$.

\item The beta parameters $(K_{SO}, \eta_{SO}$) are assigned the vague prior
$$
g(K_{SO}, \eta_{SO}) \propto \frac{1}{\eta_{SO} (1 - \eta_{SO}) (1 + K_{SO})^2}, K_{SO} > 0, 0 < \eta_{SO} < 1.
$$
\end{itemize}
Similarly, we represent a belief in exchangeability of the home run probabilities in ${\bf p_{HR}}$ by assigning a similar two-stage prior with unknown hyperparameters $K_{HR}$, and $\eta_{HR}$.  Likewise, an exchangeable prior on the hit-in-play probabilities ${\bf p_{HIP}}$ is assigned with hyperparameters $K_{HIP}$, and $\eta_{HIP}$.
 
 \subsection{The Posterior}
 
We saw that the likelihood function factors into independent components corresponding to the SO, HR, and HIP data.  Since the prior distributions of the probability vectors ${\bf p_{SO}}$, ${\bf p_{HR}}$, and ${\bf p_{HIP}}$ are independent, it follows that these probability vectors also have independent posterior distributions. We summarize standard results about the posterior of the vector of strikeout probabilities $\bf p_{SO}$ with the understanding that similar results follow for the other two vectors.

The posterior distribution of the vector $\bf p_{SO}$ can be represented by the product
$$
g({\bf p_{SO}} | {\rm data}) = g({\bf p_{SO}} | K_{SO}, \eta_{SO}, {\rm data}) \times 
g(K_{SO}, \eta_{SO} | {\rm data}) ,
$$
where $g(K_{SO}, \eta_{SO} | {\rm data})$ is the posterior distribution of the parameters of the random effects distribution, and $g({\bf p_{SO}} | K_{SO}, \eta_{SO}, {\rm data})$ is the posterior of the probabilities conditional on the random effects distribution parameters.  We discuss each distribution in turn.

\subsubsection*{Random effects distribution}

The random effects distribution $g(\eta, K | {\rm data})$ represents the ``talent curve'' of the players with respect to strikeouts, and the posterior mode $(\hat \eta_{SO}, \hat K_{SO})$ of this distribution tells us about the center and spread of this distribution.  In particular, $\hat \eta_{SO}$ represents the average strikeout rate among the players, and the estimated standard deviation
$$
SD(p) \approx \sqrt{ \frac{\hat \eta_{SO} (1 - \hat \eta_{SO})}
{\hat K_{SO} + 1}}
$$
measures the spread of this talent curve.

\subsubsection*{Probability estimates}

Given values of the parameters $\eta_{SO}$ and $K_{SO}$, the individual strikeout probabilities $p^1_{SO}$, ..., $p^N_{SO}$ have independent beta distributions where $p^j_{SO}$ is beta with shape parameters $y^j_{SO} + K_{SO} \eta_{SO}$ and $n^j - y^j_{SO} + K_{SO} (1-\eta_{SO})$, where $y_{SO}^j$ and $n^j$ represent the number of strikeouts and at-bats for the $j$th player.

Using this representation, the posterior mean of the strikeout rate for the $j$th player is given by
$$
E(p^j_{SO} | {\rm data}, \eta_{SO}, K_{SO}) = \frac{y^j_{SO} + K_{SO} \eta_{SO}}{n^j + K_{SO}}.
$$
Plugging in the posterior estimates for $\eta_{SO}$ and $K_{SO}$, we get the posterior estimate of the $j$th player striking out:
$$
\hat p^j_{SO}  = \frac{y^j_{SO} + \hat K_{SO} \hat \eta_{SO}}{n^j + \hat K_{SO}}.
$$

The same methodology was used to estimate the home run probabilities and the hit-in-play probabilities for all players -- denote the three sets of estimates as $\{\hat p^j_{SO}\}$, $\{\hat p^j_{HR}\}$, and $\{\hat p^j_{HIP}\}$, respectively.    One can use these estimates to estimate the hitting probabilities  using the representation
$$
p_H = (1 - p_{SO}) \times \left( p_{HR} + (1 - p_{HR}) \times p_{HIP}\right).
$$
For an individual estimate, by substituting the ``component'' estimates $\hat p^j_{SO}$, $\hat p^j_{HR}$, and $\hat p^j_{HIP}$ into this expression, we obtained a ``component'' estimate of $p^j_H$ that we denote by $\hat p^j_H$.

\subsection{Example:  2011 Season}

To illustrate the use of these exchangeable models, we collect hitting data for all players with at least 100 at-bats in the 2011 season.  (We used 100 AB as a minimum number of at-bats to exclude pitchers from the sample.)  Three exchangeable models are fit, one to the collection of strikeout rates \{$y_{SO}^j /  n_j$\}, one to the collection of home run rates \{$y_{HR}^j /  (n_j - y^j_{SO})$\}, and one to the collection of in-play hit rates \{$y_{HIP}^j /  (n_j - y^j_{SO} - y^j_{HR})$\}.

Table 1 displays values of the random effect parameters $\eta$ and $K$ for each of the three fits.  The average strikeout rates is about 20\%, the average home run rate (among AB removing strikeouts) is 3.6\%, and the average in-play hit rate is $30\%$.  The estimated values of $K$ are informative about the spreads of the associated probabilities.  The relatively small estimated value of $K$ for $SO$ reflects a high standard deviation, indicating a large spread in the player strikeout probabilities.  The estimated value of $K$ for home runs is also relatively small, indicating a large spread in home run probabilities.  In contrast, the estimated value of $K$ for hits in play is large, indicated that players' abilities to get in-play hits are more similar.

\begin{table}[ht]
\centering
\caption{Estimates of random effect parameters for strikeout data, home run data, and in-play hit data from the 2011 season.}
\begin{tabular}{rrrr}
  \hline
 & SO & HR & H \\ 
  \hline
$\eta$ & 0.203 & 0.0369 & 0.303 \\ 
  $K$ & 40.60 & 65.70 & 418.10 \\ 
  SD & 0.062 &  0.023 &  0.022 \\
   \hline
\end{tabular}
\end{table}

These estimates of $K$ and $\eta$ can be used to compute ``improved'' estimates at player strikeout probabilities, home run probabilities, and in-play hit probabilities, and these component estimates can be used to obtain estimates at player hit probabilities.

To illustrate these calculations, consider Carlos Beltran with the hitting statistics displayed in Table 2. 
\begin{table}[ht]
\centering
\caption{Batting data for Carlos Beltran for the 2011 season.}
\begin{tabular}{rrrrr}
  \hline
 & AB & SO & HR & H \\ 
  \hline
Count & 520 & 88 & 22 & 156 \\
   \hline
\end{tabular}
\end{table}

His three observed rates are $SO Rate = 88 / 520 = 0.169$, $HR Rate = 22 / (520 - 88) = 0.051$, and $BABIP = (156 - 22) / (520 - 88 - 22) = 0.327.$  From the fitted model,  these observed rates are shrunk or adjusted towards  average values using the formulas:
$$
\hat p_{SO} = \frac{88 + 40.60 \times 0.203}{520 + 40.60} = 0.172
$$
$$
\hat p_{HR} = \frac{22 + 65.70 \times 0.0369}{520 - 88 + 65.70} = 0.049
$$
$$
\hat p_{HIP} = \frac{156 - 22 + 418.10 \times 0.303}{520 - 88 - 22 + 418.10} = 0.315.
$$
(Note that Beltran's strikeout and home runs are slightly adjusted towards the average values due to the small estimated values of $K$.  In contrast, the consequence of the large estimated $\hat K$ is that Beltran's in-play hit rate is adjusted about half of the way towards the average value.)  Using these estimates, Beltran's hit probability is estimated to be
$$
\hat p_H =  (1 - 0.172) \times \left( 0.049 + (1 - 0.049) \times 0.315\right) = 0.289,
$$
which is smaller than his observed batting average of 156 / 520 = 0.300.  Much of this adjustment in his batting average estimate is due to the adjustment in Beltan's in-play hitting rate.

\section{Evaluation}

\subsection{A Single Exchangeable Model}

If one is primarily interested in estimating the hitting probabilities, there is a well-known simpler alternative approach based on an exchangeable model placed on the probabilities.  If the hitting probability of the $j$th player is given by $p_H^j$, then one can assume that $p_H^1, ..., p_H^N$ are distributed from a common beta curve with parameters $K$ and $\eta$, and assign $(K, \eta)$ a vague prior.  Then the posterior mean of $p_H^j$ is approximated by
$$
\tilde p_H^j = \frac{y_H^j + \tilde K \tilde \eta}{n^j + \tilde K},
$$
where the $j$th player is observed to have $y_H^j$ hits in $n^j$ at-bats and $\tilde K$ and $\tilde \eta$ are estimates from this exchangeable model.

\subsection{A Prediction Contest}

The following prediction contest is used to compare the proposed component estimates \{$\hat p^j_H$\} with the batting average estimates \{$\tilde p^j_H$\}

\begin{itemize}
\item First we collect hitting data for all players with at least 100 at-bats in both 2011 and 2012 seasons.
\item Fit the component model on the hitting data for 2011 season -- get estimates of the strikeout, home run, and in-play hit probabilities for all hitters and use these three sets of estimates to get the component estimates \{$\hat p^j_H$\}.
\item Use the hit/at-bat data for all players in the 2011 season and the single exchangeable model to compute  the estimates.

\item Use both the component estimates and the single exchangeable estimates to predict the batting averages of the players in the 2012 season.  Let $w_j$  and $m_j$ denote the number of hits and at-bats of the $j$th player in the 2012 season. Compute the root sum of squared prediction errors for both methods.
$$
S_C = \sqrt{\sum \left(\frac{w_j}{m_j} - \hat p_j \right)^2}
$$

$$
S_I = \sqrt{\sum \left(\frac{w_j}{m_j} - \tilde p_j \right)^2}
$$
The improvement in using the components estimates is $I = S_I - S_C$.  A positive value of $I$ indicates that the component estimates are providing closer predictions than the single exchangeable estimates. 
\end{itemize}

This prediction contest was repeated for each of the seasons 1963 through 2012.  Batting data for all players in seasons $y$ and $y+1$ were collected, y = 1963, ..., 2012.  The component and single exchangeable models were each fit to the data in season $y$ and the improvement in using the component method over the single exchangeable model in predicting the batting averages in season $y+1$ was computed.  Figure 2 graphs the prediction improvement as a function of the season.  It is interesting that the component estimates were not uniformly superior to the estimates from a single exchangeable model.  However the component method appears generally to be superior to the standard method, especially for seasons 1963-1980 and 1995-2012.

\begin{figure}[h]
\begin{center}
 \includegraphics[scale=0.6]{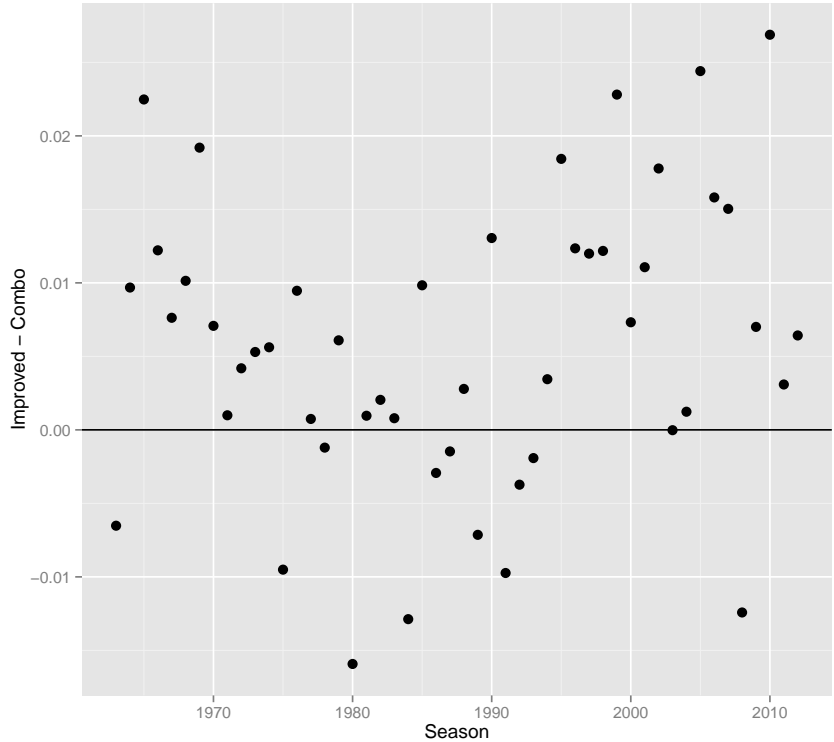}
\end{center}
\caption{Improvement in error in predicting batting averages by using the component method for each of the seasons 1963 through 2012.}
\end{figure}

\section{On-Base Percentages}

\subsection{Decomposition}

We have focused on the decomposition of an at-bat.  In a similar manner, one can decompose a plate appearance as displayed in Figure 3.  If we ignore sacrifice hits (both SH and SF), then one can express an on-base percentage as
$$
OBP \approx \frac{H + BB + HBP}{AB + BB + HBP}.
$$
If one combines walks and hit by pitches and defines the ``Walk Rate''
$$
Walk.Rate = \frac{BB + HBP}{AB + BB + HBP},
$$
then one can write
$$
OBP \approx Walk.Rate + (1 - Walk.Rate) \times BA,
$$
where $BA = H / AB$ is the batting average.

\begin{figure}[h]
\begin{center}
 \includegraphics[scale=0.4]{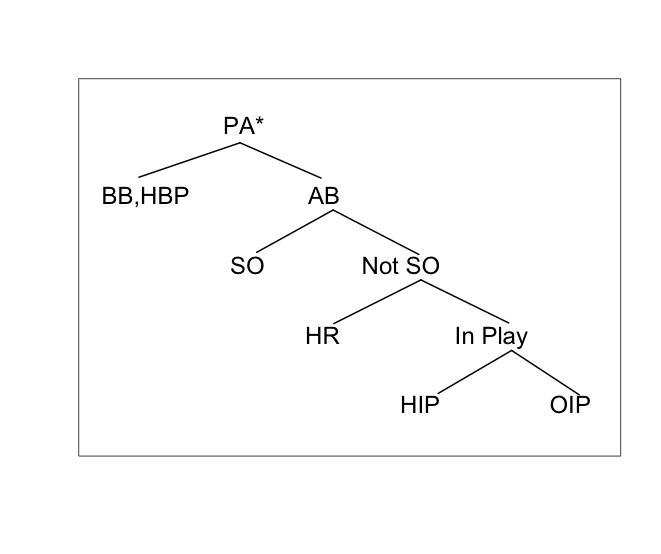}
\end{center}
\caption{Breakdown of a plate appearance.}
\end{figure}

This representation makes it clear that an OBP is basically a function of a hitter ability to draw walks, as measured by the walk rate and his batting average.  Also, following the logic of the previous section, this representation suggests that one may accurately estimate a player's on-base probability by combining separate accurate estimates of his walk probability and his hitting probability.

\subsection{Estimating On-Base Percentages}

In this setting, one can simultaneously estimate on-percentages of a group of players by separately estimating their walk probabilities and their hit probabilities.  One represents a probability that a player gets on-base $p_{OB}$ as 
$$
p_{OB} = p_{BB} \times \left(1 + (1 - p_{BB}) \times p_H\right).
$$
This suggests a method of estimating a collection of on-base probabilities.
\begin{enumerate}
\item Estimate the walk probabilities \{$p_{BB}^j$\} by use of an exchangeable model.
\item Estimate the hitting probabilities \{$p_H^j$\} by use of an exchangeable model.
\item Estimate the on-base probabilities by use of the formula
$$
\hat p^j_{OB} = \hat p^j_{BB} \times \left(1 + (1 - \hat p^j_{BB}) \times \hat p^j_H\right),
$$
where $\hat p^j_{BB}$ and $\hat p^j_H$ are estimates of the walk probability and the hit probability for the $j$th player.
\end{enumerate}

Figure 4 demonstrates the value of this method in providing better predictions.  As in the ``prediction contest'' of Section 5.2, we are interested in predicting the on-base probabilities for one season given hitting data from the previous season.  Two prediction methods are compared -- the ``single exchangeable'' method fits one exchangeable data using the on-base fractions, and the ``component'' method separately estimates the walk rates and hitting rates for the players.  One evaluates the goodness of predictions by the square root of the sum of squared prediction errors and one computes the improvement in using the component procedure over the single exchangeable method.  These methods are compared for 50 prediction contests using data from each of the seasons 1963 through 2012 to predict the on-base proportions for the following season.  As most of the points fall above the horizontal line at zero, this demonstrates that the component method generally is an improvement over the one exchangeable method.

\begin{figure}[h]
\begin{center}
 \includegraphics[scale=0.45]{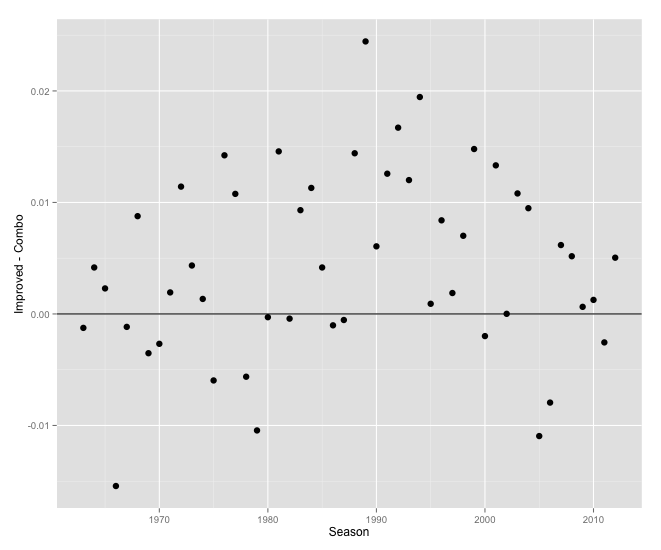}
\end{center}
\caption{Improvement in error in predicting on-base percentages by using the component method for each of the seasons 1963 through 2012.}
\end{figure}

\section{Historical Perspective of Hitting Rates}

To obtain a historical perspective of the change in hitting rates, the basic exchangeable model was  fit to rates for all batters with at least 100 AB for each of the seasons 1960 through 2012.  For each season, we estimate the mean talent  $\hat \eta$ and associated precision parameter $\hat K$ -- the associated estimated posterior standard deviation of the talent distribution is 
$$
SD(p) \approx  \sqrt{\frac{\hat \eta (1 - \hat \eta)}{\hat K + 1}}
$$

Figure 5 displays the pattern of mean strikeout rates for all batters with at least 100 AB.  Note that the average strikeout rate  among batters initially showed a decrease from 1970 through 1980 but has steadily increased until the current season.  If we performed fits of the exchangeable model for all pitchers for each season from 1960 through 2012, one would see a similar pattern in the mean strikeout rates.

Figure 6 displays the estimated standard deviations of the strikeout abilities of all batters with at least 100 AB across seasons and overlays the estimated season standard deviations of the strikeout abilities of all pitchers.  First, note that among batters, the spread of the strikeout abilities shows a similar pattern to the mean strikeout rate -- there is a decrease from 1970 to 1980 followed by a steady increase to the current day.  The spread of strikeout abilities among pitchers shows a different pattern.  The standard deviations for pitchers have steadily increased over seasons, and the spread in the talent distribution for pitchers is significantly smaller than the spread of the talents for batters.

\begin{figure}[h]
\begin{center}
\includegraphics[scale=0.25]{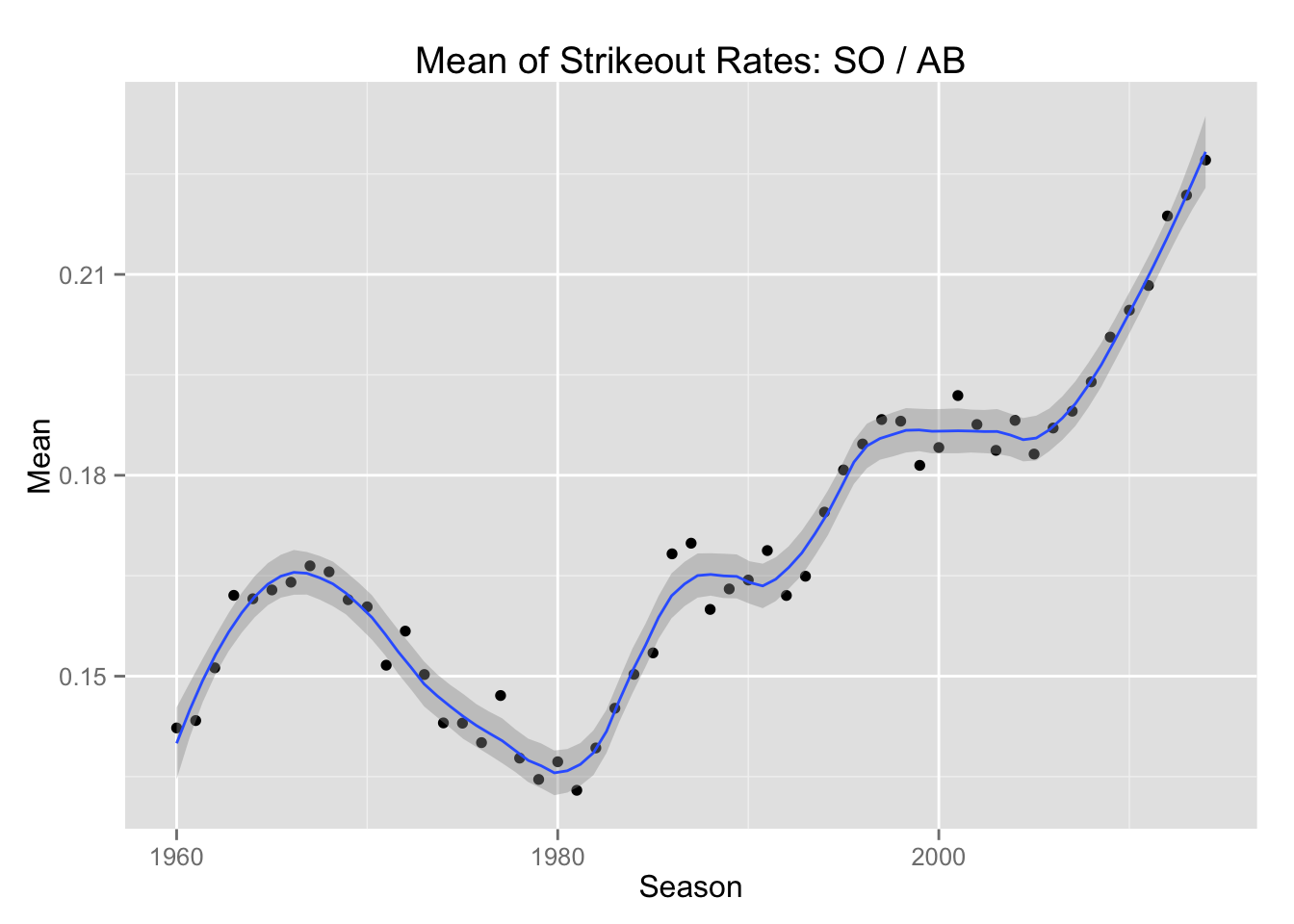}
\end{center}
\caption{Plot of mean strikeout rates for batters (at least 100 AB) for seasons 1960 through 2012.}
\end{figure}

\begin{figure}[h]
\begin{center}
\includegraphics[scale=0.55]{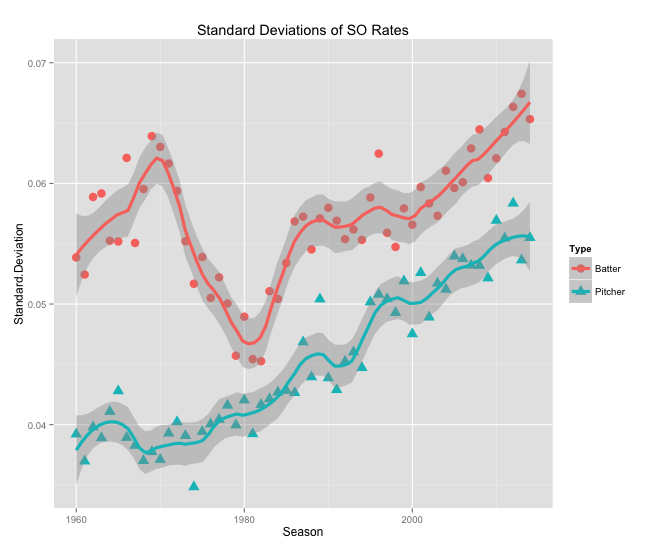}
\end{center}
\caption{Plot of standard deviation of strikeout rates for batters (at least 100 AB) and for pitchers for seasons 1960 through 2012.}
\end{figure}

\section{Career Trajectories}

\subsection{Predictive Residuals}

One way of measuring the effectiveness of a batter or a pitcher is to look at the vector of rates
$(BB.Rate, SO.Rate, HR.Rate, BABIP)$ for a particular season.  Plotting these rates over a player's career, one gains a general understanding of the strengths of the batter or pitcher and learns when these players achieved peak performances.  Albert (2002) demonstrates the value of looking at career trajectories to better understand the growth and deterioration of player's batting abilities.

The four observed rates have different averages and spreads, and as we see from Figures 5 and 6, the averages and spreads can change dramatically over different seasons.  We use residuals from the predictive distribution to standardize these rates.  Let $y$ denote the number of successes in $n$ opportunities for a player in a particular season and suppose the underlying probabilities of the players follow a beta curve with mean $\eta$ and precision $K$.
The predictive density of the rate $y/n$ has mean $\eta$ and standard deviation 
$$
SD(y/n) = \sqrt{  \eta (1 -  \eta) \left(\frac{1}{n} + \frac{1}{K + 1}\right)}.
$$

When  the exchangeable model is fit, one obtains estimates of the random effects parameters $\hat \eta$ and $\hat K$, and obtains an estimate of the standard deviation  $\widehat{SD(y/n)}$.   Define the standardized residual
$$
z = \frac{y / n - \hat \eta}{\widehat{SD(y/n)}}.
$$
In the following plots of the standardized residuals of the walk/hit-by-pitch rates, strikeout rates, home run rates, and hit-in-play rates will be displayed to show special strengths of hitters and pitchers.

\subsection{Batter Trajectories}

The graphs of the standardized rates are displayed for the careers of Mickey Mantle in Figure 7 and Ichiro Suzuki in Figure 8.  Looking at the four graphs of Figure 7 in a clockwise manner from the upper-left, one sees
\begin{itemize}
\item Mantle drew many walks/HBP and his walk/HBP rate actually increased during his career.
\item Mantle had an above-average strikeout rate.
\item His home run rate hit a peak during the middle of his career.
\item His in-play hit rate decreased towards the end of his career.
\end{itemize}
In contrast, by looking at Figure 8, one sees that Suzuki had consistent low walk/HBP, strikeout, and home run rates throughout his career.  He was especially good in his hit-in-play rate, although there was much variability in these rates and showed a decrease towards the end of his career.

\begin{figure}[h]
\begin{center}
\includegraphics[scale=0.45]{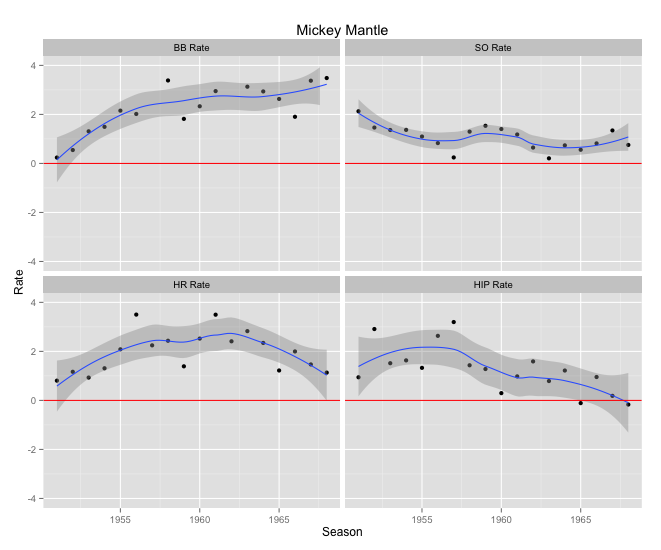}
\end{center}
\caption{Standardized residuals of the four rates for Mickey Mantle.}
\end{figure}

\begin{figure}[h]
\begin{center}
\includegraphics[scale=0.45]{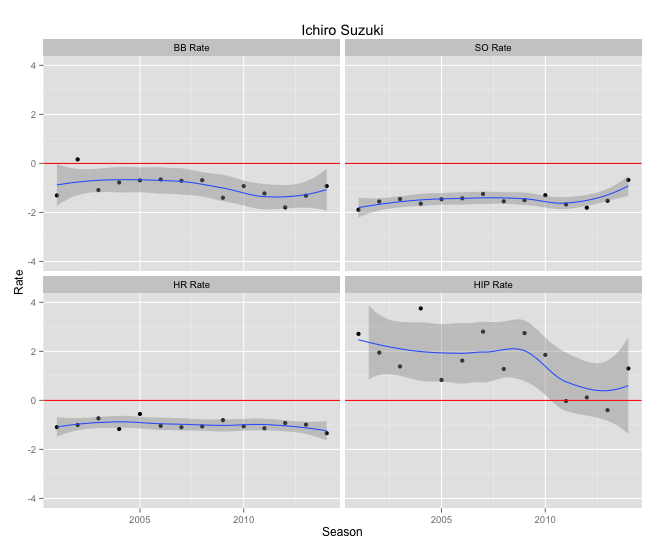}
\end{center}
\caption{Standardized residuals of the four rates for Ichiro Suzuki.}
\end{figure}

\subsection{Pitcher Trajectories}

These displays of standardized rates are also helpful for understanding the strengths of pitchers in the history of baseball.  Figures 9 and 10 display the standardized rates for the Hall of Fame pitchers Greg Maddux and Steve Carlton.  Maddux was famous for his low walk rate and generally low ERA.  Looking at the trajectories of his rates in Figure 9, one sees that Maddux's best walk rates occurred during the last half of his career.  His best strikeout rate, home run rate, and HIP rate occurred about 1995 and all three of these rates deteriorated from 1995 until his retirement in  2008.  In contrast, one sees from Figure 10 that Carlton had a slightly below average walk rate and a high strikeout rate during his career.  Since all of these rates significantly deteriorated towards the end of his career, perhaps Carlton should have retired a few years earlier.  Based on these graphs, Carlton's peak season in terms of performance was about 1980, the season when the Phillies won the World Series.

\begin{figure}[h]
\begin{center}
\includegraphics[scale=0.45]{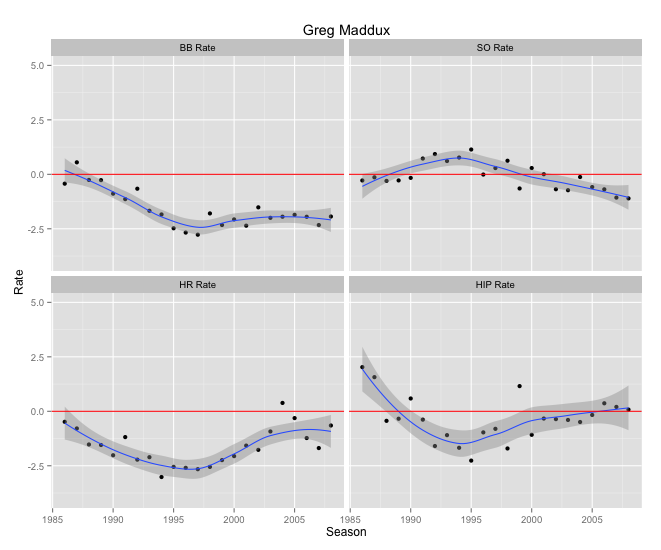}
\end{center}
\caption{Standardized residuals of the four rates for Greg Maddux.}
\end{figure}

\begin{figure}[h]
\begin{center}
\includegraphics[scale=0.45]{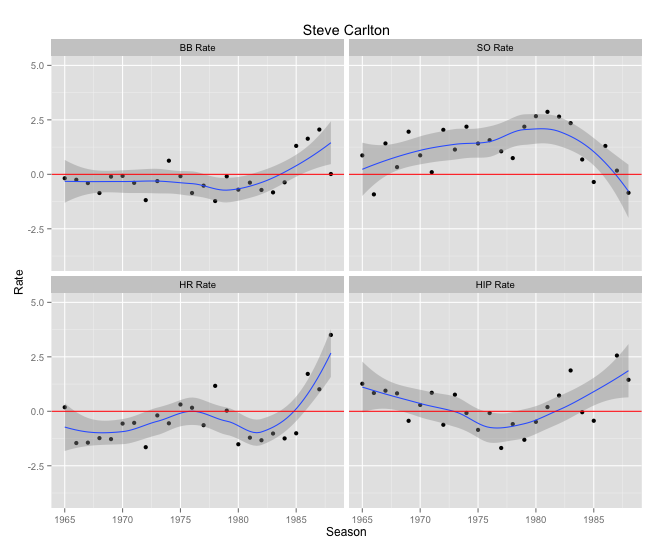}
\end{center}
\caption{Standardized residuals of the four rates for Steve Carlton.}
\end{figure}

\section{FIP Measures}

\subsection{Introduction}

Recently, there has been an increased emphasis on the use of fielding-independent-performance (FIP) measures of pitchers.  The idea is to construct a measure based on the outcomes such as walks, hit-by-pitches, strikeouts, and home runs that a pitcher  directly controls.  The usual definition of FIP is given by
$$
FIP = \frac{13 HR + 3 (BB + HBP) - 2 SO}{IP} + constant,
$$
where $HR$, $BB$, $HBP$, and $SO$ are the counts of these different events, $IP$ is the innings pitched, and $constant$ is a constant defined to ensure that the average FIP is approximately equal to the league ERA.

Although $FIP$ is defined in terms of counts, it is straightforward to write it as a function of the four rates $SO.Rate$,  $HR.Rate$, $BABIP$, and $Walk.Rate'$.   Let $BFP$ denote the count of batters faced, then

\begin{eqnarray*}
HR & = & BFP (1 - Walk.Rate) (1 - SO.Rate) HR.Rate \\
BB + HBP &=& BFP \times Walk.Rate \\
SO &=& BFP (1 - Walk.Rate) SO.Rate \\
IP &=& \frac{1}{3} BFP (1 - Walk.Rate)[SO.Rate \\ 
    & & + (1 - SO.Rate) (1 - HR.Rate)(1- BABIP)]
\end{eqnarray*}

Substituting these expressions into the formula and ignoring the constant term, the $FIP$ measure is expressed solely in terms of these four rates.  Although on face value, the $FIP$ measure seems to depend on the sample size (the number of batters faced), the value of $BFP$ cancels out in the substitution.

\subsection{Estimation of FIP Ability}

All of the observed rates are estimates of the underlying probabilities of those events.  If we take the expression of $FIP$, ignoring the constant, and replace the rates with probabilities, we get an expression for a pitcher's $FIP$ ability denoted by $\mu_{FIP}$:
$$
\mu_{FIP} = \frac{39 (1 - p_{BB})(1 - p_{SO}) p_{HR} + 9 p_{BB} - 6 (1 - p_{BB}) p_{SO}}
{(1 - p_{BB}) (p_{SO} + (1 - p_{SO}) (1 - p_{HR}) (1 - p_{HIP}))}.
$$

Using data for a single season, we can use separate exchangeable models to estimate the walk probabilities \{$p_{BB}^j$\}, the strikeout probabilities \{$p_{SO}^j$\}, the home run probabilities \{$p_{HR}^j$\}, and the hit-in-play probabilities \{$p_{HIP}^j$\} for all pitchers.  If we substitute the probability estimates into the $\mu_{FIP}$ formula, we get new estimates at the observed $FIP$ measures for all pitchers in a particular season.

\subsection{Performance}

Based on our earlier work, one would anticipate that our new estimates of $FIP$ ability would be superior  to usual estimates in predicting the $FIP$ values of the pitchers in the following season.  As in our evaluation of the performance of the improved batting probabilities, the new estimates can be compared with exchangeable estimates based on the standard representation of the $FIP$ statistic.

For a given pitcher, suppose one collects the measurement $13 HR + 3(BB + HBP) - 2 SO$ for each inning pitched.  If the pitcher pitches for $N = IP$ innings, then the measurements can be denoted by $Y_1, ..., Y_N$ and the $FIP$ statistic is simply the sample mean $FIP = \bar Y$.  It is reasonable to assume that $\bar Y$ is normal with mean $\mu_{FIP}$ and variance $\sigma^2 / N$, where $\sigma$ reflects the variability of the values of $Y_j$ within innings. 

Based on this representation, one can estimate the $FIP$ abilities \{$\mu^j_{FIP}$\} by use of an exchangeable model where the abilities are assigned a normal curve with mean $\mu$ and standard deviation $\tau$, and a vague prior is assigned to $(\mu, \tau)$.    By fitting this model, one shrinks the observed $FIP$ values for the pitchers towards an average value.

Again a prediction experiment is used to predict the $FIP$ values for all pitchers from a season given these measures from the previous season.   The ``standard'' method predicts the $FIP$ values using the single exchangeable model, and the ``component'' method first separately estimates the four sets of probabilities with exchangeable models, and then substitutes these estimates in the formula to obtain $FIP$ predictions.  As might be expected, the component method results in a smaller prediction error for practically all of the seasons of the study.
This again demonstrates the value of this ``divide and conquer'' approach to obtain superior estimates of pitcher characteristics that are functions of the underlying probabilities. 

\section{Concluding Comments}

In the sabermetrics literature, the regression effect is well known;  to predict a batter's hitting rate for a given season, one takes one's previous season's hitting average and move this estimate towards an average.  This paper extends this approach to estimating a batting measure that is a function of different rates.  Apply the random effects model to get accurate estimates at the component rates for all players, and then substitute these estimates into the function to get improved predictions of the batting measures.  This approach was  easy to apply for the batting probability and on-base probabilities situations due to the convenient factorization of the likelihood and use of independent exchangeable prior distributions.

The choice of a single beta random effects curve was chosen for convenience due to attractive analytical features, but this ``component'' approach can be used for any choice of random effects model.  For example, one may wish to use covariates in modeling the probabilities that hitters get a hit on balls put in play.  If $p^j_{HIP}$ is the probability that the $j$th player gets a hit, then one could assume that 
$p^1_{HIP}, ..., p^N_{HIP}$ are independent from beta$(\eta^1, K), ..., $beta$(\eta^N, K)$ distributions where the prior means satisfy the logistic model
$$
\log \left(\frac{\eta^j}{1- \eta^j}\right)= \beta_0 + \beta_1 x^j,
$$
where $x^j$ is a relevant predictor such as the speed of the ball off the bat.  As before, the prior parameters $(\beta_0, \beta_1, K)$ would be assigned a weakly informative prior to complete the model.

The FIP measure was motivated from the basic observation that a team defense, not just a pitcher, prevents runs, and one wishes to devise alternative measures that isolate a pitcher's effectiveness.  In a similar fashion, the goal here is to isolate the different components of a hitter's effectiveness.  These component estimates are useful by themselves, but they are also helpful in estimating ensemble measures of ability such as the probability of getting on base.


\begin{thebibliography}{99}

\bibitem{albert2002}
Albert, J. (2002), ``Smoothing career trajectories of baseball hitters,'' Technical Report, {\tt http://bayes.bgsu.edu}.

\bibitem{albert2004}
Albert, J. (2004), ``A batting average:  does it represent ability or luck?'', Technical Report, {\tt http://bayes.bgsu.edu}

\bibitem{albert2006}
Albert, J. (2006),  ``Pitching statistics, talent and luck, and the best strikeout seasons of all-time.'' {\it Journal of Quantitative Analysis in Sports},  2, issue 1.

\bibitem{baumer}
Baumer, B. (2008), ``Why on-base percentage is a better indicator of future performance than batting average: an algebraic proof,''  {\it Journal of Quantitative Analysis in Sports},  3, issue 2.

\bibitem{bickel}
Bickel, E. (2004), ``Why It’s so hard to hit .400'.'' {\it Baseball Research Journal}, 32, 15-
21.

\bibitem{bickelstotz}
Bickel, E. and Stotz, D. (2003), ``Batting average by count and pitch type,” 
{\it Baseball Research Journal}, 31, 29-34.

\bibitem{brown}
Brown, L. (2008), ``In-season prediction of batting averages: a field test of
empirical Bayes and Bayes methodologies,  {\it The Annals of Applied
Statistics,}  2(1), 113-152.

\bibitem{efronmorris}
Efron, B., and  Morris, C. (1975), ``Data analysis using Stein's estimator and its generalizations." 
{\it Journal of the American Statistical Association} 70,  311-319.

\bibitem{mccracken}
McCracken, V. (2001), ``Pitching and defense,''  {\it Baseball Prospectus}
({\tt http://www.baseballprospectus.com}).

\bibitem{mcshane}
McShane, B., Braunstein, A., Piette, J. and Jensen, S. (2011), ``A hierarchical Bayesian variable selection approach to Major League Baseball hitting metrics,'' {\it Journal of Quantitative Analysis in Sports}, 7, issue 2.

\bibitem{Neal}
Neal, D., Tan, J., Hao, F. and Wu, S. (2010), ``Simply better:  using regression models to estimate Major League batting averages,''  {\it Journal of Quantitative Analysis in Sports},  6, issue 3.

\bibitem{Null}
Null, B. (2009), ``Modeling baseball player ability with a nested Dirichlet distribution,''  {\it Journal of Quantitative Analysis in Sports},  5, issue 2.

\bibitem{Piette}
Piette, J. and Jensen, S. (2012), ``Estimating fielding ability in baseball players over time,'' {\it Journal of Quantitative Analysis in Sports,} 8, issue 3.

\bibitem{Piette2}
Piette, J., Braunstein, McShane, and Jensen (2010), ``A point-mass mixture random effects model for pitching metrics,''  {\it Journal of Quantitative Analysis in Sports},  6, issue 3.

\bibitem{Tango}
Tango, T., Lichtman, M. and Dolphin, A. (2007), {\it The Book:  Playing the Percentages in Baseball'}, Potomac Books.


\end{thebibliography}
\end{document}